\documentstyle[11pt,aaspp4,flushrt]{article}

\slugcomment{\it to appear in The Astrophysical Journal Letters}
\lefthead{Stern et al.}
\righthead{Quasar at $z = 5.50$}

\def\eg{{e.g.,}}

\def\rd300{RD~J030117+002025}

\def\deg{\ifmmode {^{\circ}}\else {$^\circ$}\fi}

\def\secper{\ifmmode \rlap.{^{s}}\else $\rlap{.}{^{s}} $\fi}

\def\kms{\ifmmode {\rm\,km\,s^{-1}}\else
    ${\rm\,km\,s^{-1}}$\fi}
\def\kmsMpc{\ifmmode {\rm\,km\,s^{-1}\,Mpc^{-1}}\else
    ${\rm\,km\,s^{-1}\,Mpc^{-1}}$\fi}
\def\ergAcm2{\ifmmode {\rm\,ergs\,cm^{-2}\,{\rm \AA}^{-1}}\else
    ${\rm\,ergs\,cm^{-2}\,\AA^{-1}}$\fi}
\def\ergcm2s{\ifmmode {\rm\,ergs\,cm^{-2}\,s^{-1}}\else
    ${\rm\,ergs\,cm^{-2}\,s^{-1}}$\fi}
\def\ergsHz{\ifmmode {\rm\,ergs\,s^{-1}\,Hz^{-1}}\else
    ${\rm\,ergs\,s^{-1}\,Hz^{-1}}$\fi}
\def\ergs{\ifmmode {\rm\,ergs\,s^{-1}}\else
    ${\rm\,ergs\,s^{-1}}$\fi}
\def\ergsA{\ifmmode {\rm\,ergs\,s^{-1}\,\AA^{-1}}\else
    ${\rm\,ergs\,s^{-1}\,\AA^{-1}}$\fi}
\def\WHz{\ifmmode {\rm\,W\,Hz^{-1}}\else
    ${\rm\,W\,Hz^{-1}}$\fi}
\def\WHzsr{\ifmmode {\rm\,W\,Hz^{-1}\,sr^{-1}}\else
    ${\rm\,W\,Hz^{-1}\,sr^{-1}}$\fi}
\def\ergscmHz{\ifmmode {\rm\,ergs\,cm^{-2}\,Hz^{-1}}\else
    ${\rm\,ergs\,cm^{-2}\,Hz^{-1}}$\fi}

\def\spose#1{\hbox to 0pt{#1\hss}}
\def\simlt{\mathrel{\spose{\lower 3pt\hbox{$\mathchar"218$}}
     \raise 2.0pt\hbox{$\mathchar"13C$}}}
\def\simgt{\mathrel{\spose{\lower 3pt\hbox{$\mathchar"218$}}
     \raise 2.0pt\hbox{$\mathchar"13E$}}}

\def\lyb{Ly$\beta$}

\def\lya{Ly$\alpha$}
\def\nv{\ion{N}{5} $\lambda$1240}

\def\siivoiv{\ion{Si}{4}/\ion{O}{4}] $\lambda$1403}

\begin{document}

\title{Discovery of a Color-Selected Quasar at $z =
5.50$\altaffilmark{1}}

\author{Daniel Stern\altaffilmark{2}, Hyron Spinrad\altaffilmark{3},
Peter Eisenhardt\altaffilmark{2}, \\ Andrew J.~Bunker\altaffilmark{4},
Steve Dawson\altaffilmark{3}, S.~A.~Stanford\altaffilmark{5},
\& Richard Elston\altaffilmark{6}}

\altaffiltext{1}{Based on observations at the W.M. Keck Observatory,
Kitt Peak National Observatory, and Palomar Observatory.  Keck
Observatory is operated as a scientific partnership among the
University of California, the California Institute of Technology, and
the National Aeronautics and Space Administration.  The Observatory was
made possible by the generous financial support of the W.M. Keck
Foundation.}

\altaffiltext{2}{Jet Propulsion Laboratory, California
Institute of Technology, Mail Stop 169-327, Pasadena, CA 91109}

\altaffiltext{3}{Department of Astronomy, University of California,
Berkeley, CA 94720}

\altaffiltext{4}{Institute of Astronomy, Madingley Road,
Cambridge, CB3 OHA, England}

\altaffiltext{5}{Physics Department, University of California, Davis,
CA 95616, and Lawrence Livermore National Laboratory}

\altaffiltext{6}{Department of Astronomy, The University of Florida,
P.O. Box 112055, Gainesville, FL 32611}

\begin{abstract}

We present observations of \rd300, a quasar at $z = 5.50$ discovered
from deep, multi-color, ground-based observations covering 74
arcmin$^2$.  This is the most distant quasar or AGN currently known.
The object was targeted as an $R$-band dropout, with $R_{\rm AB} >
26.3$ (3$\sigma$ limit in a 3\arcsec\ diameter region), $I_{\rm AB} =
23.8$, and $z_{\rm AB} = 23.4$.  The Keck/LRIS spectrum shows broad
\lya/\nv\ emission and sharp absorption decrements from the
highly-redshifted hydrogen forests.  The fractional continuum
depression due to the \lya\ forest is $D_A = 0.90$.  \rd300\ is
the least luminous, high-redshift quasar known ($M_B \approx -22.7$).

\end{abstract}


\section{Introduction}

The past few years have witnessed a watershed in our direct
observations of the high-redshift Universe.  A decade ago, only a
handful of galaxies were identified past a redshift of 3.  These
sources represented the rare beast in the cosmos:  high-redshift radio
galaxies, or galaxies associated with extremely distant, luminous
quasars.  New techniques and instruments allow us now to 
routinely identify normal, star-forming galaxies at these same epochs
\markcite{Steidel:96a}(\eg Steidel {et~al.} 1996).  \markcite{Stern:99e}Stern \& Spinrad (1999) review modern search
techniques for distant galaxies.  Improved computing power and
ambitious, large-area surveys also have pushed the frontier of distant
quasar studies \markcite{Djorgovski:99, Fan:99}(\eg Djorgovski {et~al.} 1999; Fan {et~al.} 1999).  Now we are 
regularly identifying objects which have collapsed only $\sim 1$ Gyr
after the Big Bang.  Such observations tell us about the earliest
phases of galaxy and structure formation and probe the conditions of
the early Universe.

The identification of high-redshift quasars is especially important
for several reasons.  First, quasars at early cosmic epoch require the
rapid formation of a supermassive black hole.  Assuming black holes are
not primordial, this requires the condensation of a large cloud of
hydrogen, presumably embedded within a dark matter halo.  Additionally,
the presence of metal lines in quasars demand a previous generation of
stars (two generations for nitrogen).  High-redshift quasars thus
constrain models of galaxy and structure formation
\markcite{Loeb:93, Eisenstein:95}(\eg Loeb 1993; Eisenstein \& Loeb 1995).  Also, quasars provide valuable
probes of the intervening intergalactic medium \markcite{Rauch:98}(\eg Rauch 1998)
and the intergalactic ionizing background.  For example, the absence of
a smooth depression in quasar continua short-ward of the \lya\ emission
strongly constrains the amount of neutral hydrogen in the intergalactic
medium \markcite{Gunn:65}(Gunn \& Peterson 1965).  \markcite{Songaila:99}Songaila {et~al.} (1999) find no Gunn-Peterson
trough out to redshift 5 from deep spectroscopic observations of
SDSSp~J033829.31+002156.3 at $z = 5.00$ \markcite{Fan:99}(Fan {et~al.} 1999).

In this Letter, we report the discovery of a quasar at $z = 5.50$, the
most distant quasar identified to date.  The previous  most distant
quasar was SDSSp~J120441.73$-$002149.6 at $z = 5.03$ \markcite{Fan:00}(Fan {et~al.} 2000).
At $z = 5.50$, an $H_0 = 50 \kmsMpc, \Lambda = 0, \Omega = 1$ (0.1)
universe is 790 Myr (1.51 Gyr) old, corresponding to a look-back time
of 94.0\%\ (90.9\%) of the age of the universe.  For the lambda
cosmology supported by recent studies of distant supernovae, $H_0 = 65
\kmsMpc, \Lambda = 0.7, \Omega_{\rm m} = 0.3$, the universe is 1.11 Gyr
old at $z=5.50$, corresponding to a look-back time of 92.4\%\ of the
age of the universe.

\section{Observations and Target Selection}

\rd300\ was identified from deep $RIz$-band imaging using a slightly
redder version of the `dropout' color selection techniques which have
proved successful at identifying high-redshift galaxies
\markcite{Steidel:96a, Dey:98, Spinrad:98}(\eg Steidel {et~al.} 1996; Dey {et~al.} 1998; Spinrad {et~al.} 1998) and quasars
\markcite{Kennefick:95, Djorgovski:99, Fan:99, Stern:00b}(\eg Kennefick, Djorgovski, \&  de~Calvalho 1995; Djorgovski {et~al.} 1999; Fan {et~al.} 1999; Stern {et~al.} 2000b).  The
selection criteria rely upon absorption from the \lya\ and
\lyb\ forests attenuating the rest-frame ultraviolet continua.  At $z
\simeq 5$, such objects will disappear from the $R$-band.  Long-ward of
the redshifted \lya, both quasars and star-forming galaxies display
relatively flat (in $f_\nu$) continua.  In concept, our survey is
similar to established quasar surveys relying upon the digitized
Palomar Sky Survey \markcite{Djorgovski:99}(\eg Djorgovski {et~al.} 1999) or Sloan Digital Sky
Survey \markcite{Fan:99}(\eg Fan {et~al.} 1999).  In practice, we probe a much smaller
area of sky (eventually a few $\times$ 100 arcmin$^2$) to much fainter
magnitudes.  Our survey is designed to study the high-redshift,
``normal'' galaxy population, but is also sensitive to (low-luminosity)
high-redshift quasars.

The $Iz$ imaging was obtained using the Kitt Peak National Observatory
150\arcsec\  Mayall telescope  with its Prime Focus CCD imager (PFCCD)
equipped with a thinned AR coated $2048 \times 2048$ Tektronics CCD.
This configuration gives a $14.3 \times 14.3$ arcminute field of view
with 0\farcs43 pixels. The CCD was operated using ``short scan'', where
the CCD was mechanically displaced while its charge is shifted in the
opposite direction to reduce fringing at $I$ and $z$ to very low
levels.  Two hours of Mould $I$-band ($\lambda_c = 8200$ \AA; $\Delta
\lambda = 1820$ \AA) data were obtained on UT 1995 August 31.  The
$z$-band (RG850, long-pass filter) data were obtained during UT 1997
November $4 - 6$, and the summed image represents 3.3 hours of
integration.  The combined, processed $I$ and $z$ images reach limiting
magnitudes of 25.7 and 24.8 mag, respectively (3$\sigma$ limits in
3\arcsec\ diameter apertures; AB magnitudes are used throughout this
Letter) and have 0\farcs9 and 1\farcs2 seeing, respectively.  These
images comprise one field in the $BRIzJK$ \markcite{Elston:00}Elston, Eisenhardt, \& Stanford (2000) field
galaxy survey.

On UT 1999 November $11 - 12$, we used the COSMIC camera
\markcite{Kells:98}(Kells {et~al.} 1998) on the 200\arcsec\ Hale telescope at Palomar
Observatory to obtain extremely deep (4.4~hour) Kron-Cousins $R$-band
($\lambda_c = 6200$ \AA; $\Delta \lambda = 800$ \AA) imaging of the same
field, with the purpose of identifying high-redshift candidates.
COSMIC uses a $2048 \times 2048$ pixel SITe (formerly Tektronix)
thinned CCD with 0\farcs2846 pixels, yielding a $9.7 \times 9.7$ arcmin
field-of-view.  Our combined, processed $R$-band image has 1\farcs2
seeing and reaches a depth of 26.3 mag (3$\sigma$ limit in
3\arcsec\ diameter aperture).

\begin{figure}[!t]
\begin{center}
\plotfiddle{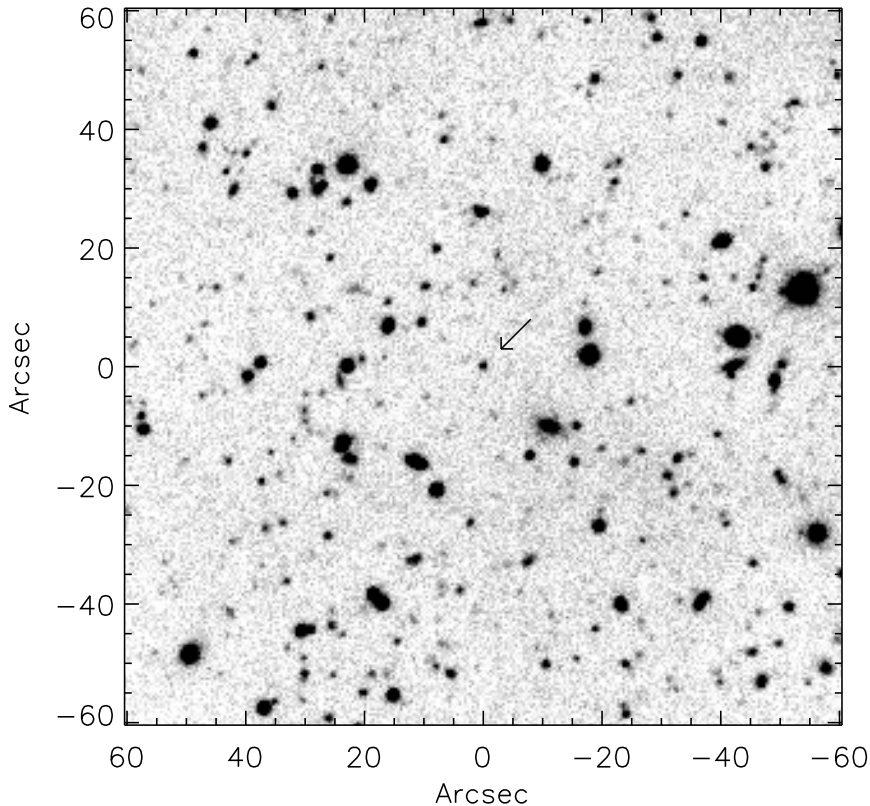}{3.7in}{0}{70}{70}{-225}{-55}
\end{center}

\caption{Finding chart for \rd300, a quasar at $z = 5.50$, from the
KPNO $I$-band imaging.  The field is 2\arcmin $\times$ 2\arcmin,
centered at $\alpha = 03^{\rm h} 01^{\rm m} 17.01^{\rm s}, \delta =
+00\deg 20\arcmin 25\farcs96$ (J2000).  North is at the top and east is
to the left.  The quasar is unresolved in this 0\farcs9 seeing image.}

\label{fig_findingchart}
\end{figure}

High-redshift candidates for spectroscopy, designated RD for $R$-drop,
were identified on the basis of a strong $R-I$ color index and
relatively flat $I-z$ color.  No morphological criteria were
implemented as the primary goal of this program is to study ``normal,''
star-forming galaxies at high redshift.  Candidates were then screened
by eye, yielding a total of six good targets over the central 74
arcmin$^2$ field.  Fig.~\ref{fig_findingchart} presents a finding chart
for \rd300, the brightest of our candidates and the subject of this
Letter.  Other candidates will be discussed in a future publication.


We obtained spectra of several $R$-band dropouts through 1\farcs5 wide,
13\arcsec\ $-$ 44\arcsec\ long slitlets using the Low-Resolution
Imaging Spectrometer \markcite{Oke:95}(LRIS; Oke {et~al.} 1995) on the Keck~II telescope
on UT 2000 January 10 and 11.  Observations were obtained at a position
angle of $-$111.6\deg\ (east of north) with the 150 lines mm$^{-1}$
grating ($\lambda_{\rm blaze} = 7500$ \AA; $\Delta \lambda_{\rm FWHM}
\approx 17$ \AA).  The spectra sample the wavelength range 4000 \AA\ to
1$\mu$m.  Seeing was $\approx$ 1\farcs1 during both nights and
conditions were photometric.  We performed $\approx$ 3\arcsec\ spatial
offsets between each 1800~s exposure in order to facilitate removal of
fringing at long wavelength ($\lambda \simgt 7200$ \AA).

\begin{figure}[!t]
\begin{center}
\plotfiddle{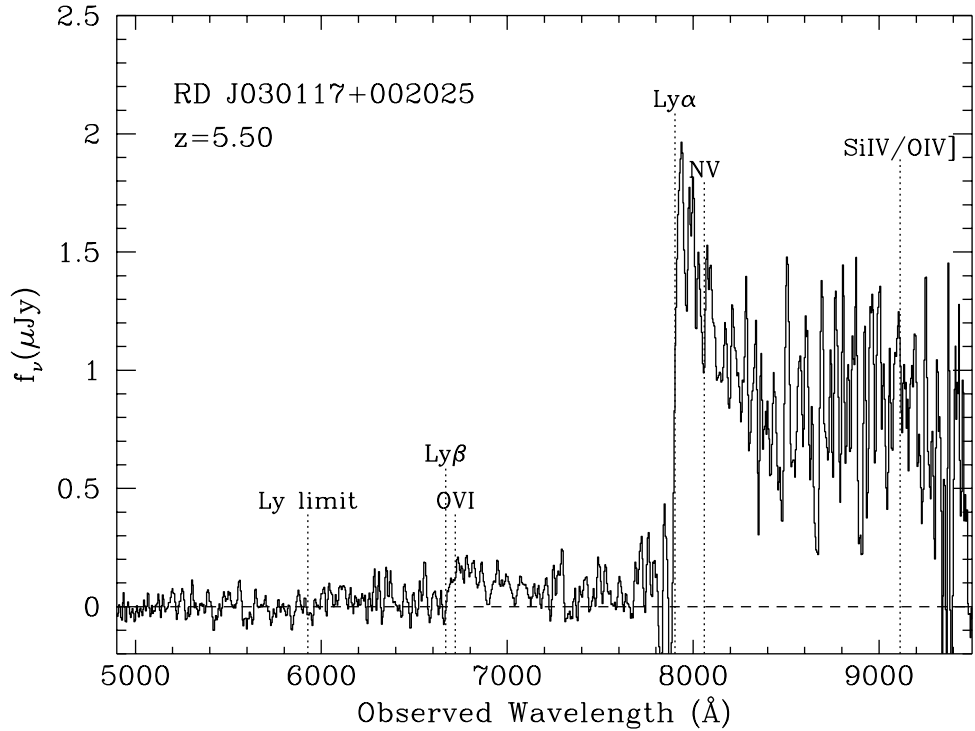}{3.4in}{-90}{50}{50}{-200}{285}
\end{center}

\caption{Spectrum of \rd300\ at $z = 5.50$, obtained with LRIS on the
Keck~II telescope.  The total exposure time is 4.5~hours, and the
spectrum was extracted using a 1\farcs5 $\times$ 1\farcs5 aperture.
The spectrum has been smoothed using a 15 \AA\ boxcar filter.  Vertical
dotted lines indicate the expected wavelength of common spectroscopic
features in quasars; not all are detected.}

\label{fig_spectrum}
\end{figure}

All data reductions were performed using IRAF and followed standard
slit spectroscopy procedures.  We calculated the dispersion using a
HgNeArKr lamp spectrum observed immediately subsequent to the science
observations (RMS variations of 0.6 \AA), and employed telluric
emission lines to adjust the wavelength zero-point.  The spectra were
flux-calibrated using observations of Feige~67 and Feige~110
\markcite{Massey:90}(Massey \& Gronwall 1990).  We corrected for foreground Galactic extinction
using a reddening of $E_{\rm B-V} = 0.03$ determined from the dust maps
of \markcite{Schlegel:98}Schlegel, Finkbeiner, \& Davis (1998).  The final composite spectrum of \rd300,
presented in Fig.~\ref{fig_spectrum}, represents 4.5~hours of
integration.


\section{Results and Discussion}

Though of moderate signal-to-noise ratio, the spectrum of \rd300\ has
the unambiguous signature of an extremely distant quasar.  The broad
emission with a sharp absorption at 7900 \AA\ is consistent with
\lya/\nv\ emission attenuated by the nearly opaque \lya\ forest at $z =
5.50$.  An additional discontinuity is visible at 6690 \AA, associated
with the \lyb\ forest.  The \lya\ forest absorption and poor detection of
the long-wavelength \siivoiv\ emission make centroiding on the emission
features ill-advised; the redshift is instead determined from the sharp
forest decrements.  We estimate $z = 5.50 \pm 0.02$.  This and other
properties of \rd300\ are given in Table~1.

The spectral character of \rd300\ is slightly atypical of high-redshift
quasars, though it undoubtedly resides within the diverse category of
quasars.  The \lya/\ion{N}{5} complex is unusually broad and
distinguishing the emission lines is impractical.  Many of the highest
redshift quasars share similar spectroscopic shapes, \eg\
SDSSp~J033829.31+002156.3 at $z = 5.00$, SDSSp~J021102.72$-$000910.3 at
$z = 4.90$ \markcite{Fan:99}(Fan {et~al.} 1999), and GB~1428+4217 at $z = 4.72$
\markcite{Hook:98}(Hook \& McMahon 1998).

The strong continuum absorption associated with the \lya\ forest is the
dominant spectroscopic feature of \rd300.  A robust determination of
$D_A$, the standard parameter for describing the \lya\ forest decrement
\markcite{Oke:82}(Oke \& Korycansky 1982), requires knowledge of the continuum spectral slope
long-ward of \lya.  We estimate $D_A$ by assuming the standard quasar
power law spectral index of $-0.5$ for the continuum long-ward of
\lya\ \markcite{Richstone:80, Schneider:92}(\eg Richstone \& Schmidt 1980; Schneider {et~al.} 1992), with the amplitude
determined over the wavelength interval $\lambda \lambda 8400 - 9000$
\AA\ (between the \nv\ and \siivoiv\ emission complexes).  We derive
$D_A = 0.90 \pm 0.02$.  We derive $D_B = 0.95 \pm 0.04$ for the
strength of the Ly$\beta$ forest.  The $D_A$ value is comparable to
those measured for distant galaxies in the Hubble Deep Field at similar
redshifts \markcite{Weymann:98, Spinrad:98}(Weymann {et~al.} 1998; Spinrad {et~al.} 1998) and models of the \lya\ forest
\markcite{Madau:95, Zhang:97}(Madau 1995; Zhang {et~al.} 1997).

At $z = 5.50$, the features used to describe continuum properties are
redshifted to challenging wavelengths.  We estimate $AB_{\rm 1450(1 +
z)}$ using the continuum modeled above.  Consistent with previous work
in this field, $M_B$ is calculated for an Einstein-de~Sitter universe
with $H_0 = 50 \kmsMpc, q_0 = 0.5$ and the standard quasar power
law index of $-0.5$.  We find $M_B \approx -22.7$; see
\markcite{Stern:00a}Stern {et~al.} (2000a) for details of how we calculate $M_B$.

Comparison with the 1.4~GHz FIRST survey \markcite{Becker:95}(Becker, White, \& Helfand 1995) reveals no
radio source within 30\arcsec\ of the quasar to a limiting flux density
of $f_{\rm 1.4 GHz} \simeq 1$ mJy (5$\sigma$).

How unusual is it to find a quasar as distant and luminous as \rd300\
in a $\simeq 100$ arcmin$^2$ field?  This is difficult to answer as
\rd300\ is the least luminous $z > 4$ quasar known.  The previously
known high-redshift ($z > 4$) quasars of lowest luminosity are
PC~0027+0521 \markcite{Schneider:94}($z = 4.21, M_B = -24.0$; Schneider, Schmidt, \& Gunn 1994), which
was discovered serendipitously, and the X-ray selected quasar
RX~J105225.9+571905 \markcite{Schneider:98}($z = 4.45, M_B = -23.9$; Schneider {et~al.} 1998).
These luminosities are comparable to the lower luminosity objects in
the $z \simlt 1$ Bright Quasar Survey \markcite{Schmidt:83}(Schmidt \& Green 1983).  Most
high-redshift quasar luminosity functions \markcite{Schmidt:95b}(\eg Schmidt, Schneider, \& Gunn 1995)
have been derived from samples of quasars $\approx 100$ times as
luminous as \rd300.  To calculate the expected surface density of
high-redshift, faint quasars, we follow the methodology
outlined in \markcite{Kennefick:95}Kennefick {et~al.} (1995) and \markcite{Boyle:98}Boyle \& Terlevich (1998):  we adopt the
\markcite{Boyle:91}Boyle {et~al.} (1991) $z = 2$ quasar luminosity function (for $q_0 = 0.5$),
scaled down in density using the evolution predicted by
\markcite{Schmidt:95b}Schmidt {et~al.} (1995), namely, that the quasar space density falls
off by a factor of 2.7 per unit redshift beyond $z = 3$.  The predicted
surface density of $R$-drop ($4.3 \simlt z \simlt 5.8$) quasars with
$M_B \leq -22.5$ is $\simeq 2 \times 10^{-3}$ arcmin$^{-2}$, implying
$\approx 0.15$ such quasars should have been uncovered in our survey.

It is dangerous, yet enticing, to draw conclusions from a single
object.  The discovery of the quasar \rd300\ at $z = 5.50$ in the
modest sky coverage of our survey is suggestive of less dramatic
evolution in the quasar luminosity function at faint magnitudes and
high redshift.  Such a change would have significant cosmological
implications, including changing the budget of high-energy, ionizing
photons in the early Universe.  We also note that the low
signal-to-noise ratio data are suggestive of strong hydrogen absorption
near the quasar redshift.  Is this due to a neutral hydrogen cloud near
the quasar, at odds with the proximity effect?  Or are we
seeing the first glimpses of an object radiating prior to the
reionization epoch, with neutral intergalactic hydrogen absorbing the
rest-frame UV photons?  Higher resolution, higher signal-to-noise ratio
data will be essential for answering these questions.


\acknowledgements

We are indebted to the expertise of the staffs of Kitt Peak, Palomar,
and Keck Observatories for their help in obtaining the data presented
herein, and to the efforts of Bev Oke and Judy Cohen in designing,
building, and supporting LRIS.  We especially thank Barbara Schaeffer,
Greg Wirth, and Jerome at Keck~II for their assistance during the
January 2000 observing run.  We are grateful to Carlos DeBreuck and
Richard McMahon for carefully reading the manuscript and to the
referee, Ray Weymann, for prompt and helpful comments.  Portions of
this work were carried out by the Jet Propulsion Laboratory, California
Institute of Technology, under a contract with NASA.  Portions of this
work was performed under the auspices of the U.S.  Department of Energy
by University of California Lawrence Livermore National Laboratory
under contract No.  W-7405-Eng-48.  This work has been supported by the
following grants:  NSF grant AST~95$-$28536 (HS), the Cambridge
Institute of Astronomy PPARC observational rolling grant
ref.~no.~PPA/G/O/1997/00793 (AJB), and NSF CAREER grant AST~9875448
(RE).


\eject

\begin{deluxetable}{lclc}
\tablecaption{Properties of \rd300}
\tablehead{
\colhead{Parameter} &
\colhead{Value} &
\colhead{Parameter} &
\colhead{Value}}
\startdata
$\alpha$ (J2000) & $03^{\rm h} 01^{\rm m} 17.01^{\rm s}$ & $z$ & $5.50 \pm 0.02$ \nl
$\delta $ (J2000) & $+00\deg 20\arcmin 25\farcs96$ & $AB_{\rm 1450 (1 + z)}$ (mag) & 24.1 \nl
$R_{\rm AB}$ (mag) & $> 26.3$ & $M_B$ (mag) & $-22.7$  \nl
$I_{\rm AB}$ (mag) & 23.8 & $W_{\rm Ly\alpha/NV}^{\rm obs}$ (\AA) & $\simeq 300$\nl
$z_{\rm AB}$ (mag) & 23.4 & $D_A$ & $0.90 \pm 0.02$ \nl
		   &      & $D_B$ & $0.95 \pm 0.04$ \nl
\enddata

\tablecomments{$R$-band magnitude is 3$\sigma$ limit is a 3\arcsec\
diameter aperture.}

\label{table1}
\end{deluxetable}


\begin{thebibliography}{}

\bibitem[Becker, White, \& Helfand 1995]{Becker:95}
Becker, R.~H., White, R.~L., \& Helfand, D.~J. 1995, \apj, 450, 559

\bibitem[Boyle, Jones, Shanks, Marano, Zitelli, \&  Zamorani 1991]{Boyle:91}
Boyle, B.~J., Jones, L.~R., Shanks, T., Marano, B., Zitelli, V., \& Zamorani,  G. 1991, in {\it The Space Distribution of Quasars}, ed. D.~Crampton, Vol.~21  (San Francisco: ASP Conference Series), 191

\bibitem[Boyle \& Terlevich 1998]{Boyle:98}
Boyle, B.~J. \& Terlevich, R.~J. 1998, \mnras, 293, L49

\bibitem[Dey, Spinrad, Stern, Graham, \& Chaffee 1998]{Dey:98}
Dey, A., Spinrad, H., Stern, D., Graham, J.~R., \& Chaffee, F. 1998, \apj, 498,  L93

\bibitem[Djorgovski, Gal, Odewahn, de~Calvalho,  Brunner, Longo, \& Scaramella 1999]{Djorgovski:99}
Djorgovski, S.~G., Gal, R.~R., Odewahn, S.~C., de~Calvalho, R.~R., Brunner, R.,  Longo, G., \& Scaramella, R. 1999, in {\it Wide Field Surveys in Cosmology},  ed. Y.~Mellier \& S.~Colombi (Gif sur Yvette: Editions Fronti\`eres), 89

\bibitem[Eisenstein \& Loeb 1995]{Eisenstein:95}
Eisenstein, D.~J. \& Loeb, A. 1995, \apj, 443, 11

\bibitem[Elston, Eisenhardt, \& Stanford 2000]{Elston:00}
Elston, R., Eisenhardt, P., \& Stanford, S.~A. 2000, in preparation

\bibitem[Fan {et~al.} 1999]{Fan:99}
Fan, X. {et~al.} 1999, \aj, 118, 1

\bibitem[Fan {et~al.} 2000]{Fan:00}
---. 2000, \aj, 119, 1

\bibitem[Gunn \& Peterson 1965]{Gunn:65}
Gunn, J.~E. \& Peterson, B.~A. 1965, \apj, 142, 1633

\bibitem[Hook \& McMahon 1998]{Hook:98}
Hook, I.~M. \& McMahon, R.~G. 1998, \mnras, 294, L7

\bibitem[Kells, Dressler, Sivaramakrishnan, Carr, Koch,  Epps, Hilyard, \& Pardeilhan 1998]{Kells:98}
Kells, W., Dressler, A., Sivaramakrishnan, A., Carr, D., Koch, E., Epps, H.,  Hilyard, D., \& Pardeilhan, G. 1998, \pasp, 110, 1487

\bibitem[Kennefick, Djorgovski, \&  de~Calvalho 1995]{Kennefick:95}
Kennefick, J.~D., Djorgovski, S.~G., \& de~Calvalho, R.~R. 1995, \aj, 110, 2553

\bibitem[Loeb 1993]{Loeb:93}
Loeb, A. 1993, \apj, 403, 542

\bibitem[Madau 1995]{Madau:95}
Madau, P. 1995, \apj, 441, 18

\bibitem[Massey \& Gronwall 1990]{Massey:90}
Massey, P. \& Gronwall, C. 1990, \apj, 358, 344

\bibitem[Oke \& Korycansky 1982]{Oke:82}
Oke, J.~B. \& Korycansky, D.~G. 1982, \apj, 255, 11

\bibitem[Oke {et~al.} 1995]{Oke:95}
Oke, J.~B. {et~al.} 1995, \pasp, 107, 375

\bibitem[Rauch 1998]{Rauch:98}
Rauch, M. 1998, \araa, 36, 267

\bibitem[Richstone \& Schmidt 1980]{Richstone:80}
Richstone, D.~O. \& Schmidt, M. 1980, \apj, 235, 361

\bibitem[Schlegel, Finkbeiner, \& Davis 1998]{Schlegel:98}
Schlegel, D., Finkbeiner, D., \& Davis, M. 1998, \apj, 500, 525

\bibitem[Schmidt \& Green 1983]{Schmidt:83}
Schmidt, M. \& Green, R.~F. 1983, \apj, 269, 352

\bibitem[Schmidt, Schneider, \& Gunn 1995]{Schmidt:95b}
Schmidt, M., Schneider, D.~P., \& Gunn, J.~E. 1995, \aj, 110, 68

\bibitem[Schneider, Schmidt, \& Gunn 1994]{Schneider:94}
Schneider, D.~P., Schmidt, M., \& Gunn, J.~E. 1994, \aj, 107, 1245

\bibitem[Schneider, Schmidt, Hasinger, Lehmann, Gunn,  Giacconi, Tr\"umper, \& Zamorani 1998]{Schneider:98}
Schneider, D.~P., Schmidt, M., Hasinger, G., Lehmann, I., Gunn, J.~E.,  Giacconi, R., Tr\"umper, J., \& Zamorani, G. 1998, \aj, 115, 1230

\bibitem[Schneider, van Gorkom, Schmidt, \&  Gunn 1992]{Schneider:92}
Schneider, D.~P., van Gorkom, J.~H., Schmidt, M., \& Gunn, J.~E. 1992, \aj,  103, 1451

\bibitem[Songaila, Hu, Cowie, \& McMahon 1999]{Songaila:99}
Songaila, A., Hu, E.~M., Cowie, L.~L., \& McMahon, R.~G. 1999, \apj, 525, L5

\bibitem[Spinrad, Stern, Bunker, Dey, Lanzetta, Yahil,  Pascarelle, \& Fern\`andez-Soto 1998]{Spinrad:98}
Spinrad, H., Stern, D., Bunker, A.~J., Dey, A., Lanzetta, K., Yahil, A.,  Pascarelle, S., \& Fern\`andez-Soto, A. 1998, \aj, 116, 2617

\bibitem[Steidel, Giavalisco, Dickinson, \&  Adelberger 1996]{Steidel:96a}
Steidel, C.~S., Giavalisco, M., Dickinson, M., \& Adelberger, K.~L. 1996, \aj,  112, 352

\bibitem[Stern, Djorgovski, Perley,  de~Carvalho, \& Wall 2000a]{Stern:00a}
Stern, D., Djorgovski, S.~G., Perley, R., de~Carvalho, R., \& Wall, J.  2000a, \aj, in press (April; astro-ph/0001394)

\bibitem[Stern, Odewahn, Gal, Djorgovski,  de~Carvalho, van Breugel, \& Spinrad 2000b]{Stern:00b}
Stern, D., Odewahn, S.~C., Gal, R., Djorgovski, S.~G., de~Carvalho, R., van  Breugel, W., \& Spinrad, H. 2000b, \aj, in preparation

\bibitem[Stern \& Spinrad 1999]{Stern:99e}
Stern, D. \& Spinrad, H. 1999, \pasp, 111, 1475

\bibitem[Weymann, Stern, Bunker, Spinrad, Chaffee,  Thompson, \& Storrie-Lombardi 1998]{Weymann:98}
Weymann, R., Stern, D., Bunker, A.~J., Spinrad, H., Chaffee, F., Thompson, R.,  \& Storrie-Lombardi, L. 1998, \apj, 505, L95

\bibitem[Zhang, Anninos, Norman, \& Meiksin 1997]{Zhang:97}
Zhang, Y., Anninos, P., Norman, M.~L., \& Meiksin, A. 1997, \apj, 485, 496

\end{thebibliography}
\end{document}